\documentstyle[psfig,conf_iap]{article}

\newcommand{\la}{\ \raise -2.truept\hbox{\rlap{\hbox{$\sim$}}\raise5.truept
        \hbox{$<$}\ }}
\newcommand{\ga}{\ \raise -2.truept\hbox{\rlap{\hbox{$\sim$}}\raise5.truept
        \hbox{$>$}\ }}

\newcommand{\be}{\begin{equation}}
\newcommand{\ee}{\end{equation}}
\newcommand{\ba}{\begin{eqnarray}}
\newcommand{\ea}{\end{eqnarray}}
\begin{document}
\heading{%
%
The UV Background and the Ly$\alpha$ Clouds at High Redshift\footnote[4]{ To 
appear in the Proceedings of the 13$^{\rm
th}$ IAP Colloquium: Structure and Evolution of the Intergalactic
Medium from QSO Absorption Line Systems, Eds. P. Petitjean and
S. Charlot.  }
%
} 
\par\medskip\noindent
\author{%
E. Giallongo$^{1}$, S. Cristiani$^{2}$, S. D'Odorico$^{3}$,
A. Fontana$^{1}$, S. Savaglio$^{3}$
}

\address{%
Osservatorio Astronomico di Roma, I-00040
Monteporzio, Italy }
\address{%
Dipartimento di Astronomia dell'Unversit\`a di Padova, I-35122 Padova, Italy
}
\address{%
European Southern Observatory, K.Schwarzschild-Strasse 2, D-85748
Garching bei M\"unchen, Germany }

\begin{abstract}
The estimate of the ionizing UV background through the proximity
effect analysis is discussed.  Taking into account the correct bending
of the column density distribution which appears in high resolution
high s/n data, a value of the UV background $\sim 5\times
10^{-22}~cgs$ is obtained in the redshift interval $z=1.7-4.1$ without
indication of any appreciable redshift evolution.  Inferences on the
spectral shape of the UV background between the HI and HeII edges are
provided by the study of metal line ratios like SiIV/CIV in optically
thin systems. The presence of a jump near the HeII edge seems favoured
at very high redshifts $z>3$. An increasing HeII jump with increasing
redshift can be responsible for the decreasing in the minimum
temperatures of the Ly$\alpha$ clouds with increasing $z$. New studies
of the cosmic evolution of star-forming galaxies allow the estimate of
a galaxy-dominated UV background intensity $\sim 1.3~ 10^{-21} \langle
f_{esc}\rangle ~cgs$ at $z\sim 0.5$ where $\langle f_{esc}\rangle
<20\%$ is the escape fraction of ionizing UV photons from the
galaxies. Finally it is also shown that in a UV background dominated
by QSOs and/or star-forming galaxies the cosmological baryon density
of the Ly$\alpha$ clouds decreases rapidly with cosmic time possibly
due to increasing bulk heating in the intergalactic medium with cosmic
time.

\end{abstract}
\section{Introduction}
 
The measure of the UV background is crucial in cosmology for two main
reasons: 1) it provides an estimate of the cosmological density of the
photoionized gas in the intergalactic medium which is consistent with
the observed Gunn-Peterson opacity; 2) it provides unique information
about the nature and abundances of the sources responsible for the
ionizing flux \cite{bechto87}.
The Lya statistics near and far away from each quasar provides
information about the strength of the UVB at the Lyman Limit as a
function of redshift under the hypothesis that the proximity effect we
see near each QSO is due to an increase of the photoionization in the
nearby clouds \cite{bajt88}.
The estimate of the UVB intensity at the Lyman edge through the
analysis of the proximity effect in the Lya forest depends both on
the accurate measure of the physical parameters related to the nearby
QSO (e.g. the systemic redshift of the QSO and the amount of the
ionizing QSO flux) and on the accurate description of the statistical
distribution of the Ly$\alpha$ lines both in column density and
redshift (see e.g. \cite{bechto94}).
There are several results about the intensity of the UVB at $z\simeq
3$ measured by the analysis of the proximity effect and the values
range in the interval $3<J_{-22}<30$ in units of $10^{-22}$ erg
s$^{-1}$ cm$^{-2}$ Hz$^{-1}$ sr$^{-1}$. However the results have been
obtained using data of very different resolution and s/n ratio and
have been analyzed using different statistical procedures.

\section {The UVB and the Ly$\alpha$ column density distribution}

There are obvious systematic effects that should be stressed in this
context. It is clear that the resolution of the spectra used for the
Lya statistics plays an important role. Indeed the proximity effect
strongly depends on the shape (i.e. the slope for a power-law
distribution) of the column density distribution. In the case of data
taken at a resolution $\geq 1$ {\AA} we can only measure equivalent
widths and an assumption must be made on the actual column density
distribution. Nevertheless, we progressively underestimate the actual
number of lines when we go away from the emission redshift since line
blending becomes stronger where the number of lines increases. As a
consequence, the observed decrement we see in the number of lines when
we approach the emission redshift will be lower in low resolution
data. For a given QSO flux, a lower decrement implies a higher UVB.

The decrement of the line number density due to the decrement of the
observed column densities is a function of the shape of the column
density distribution which is affected by the spectral resolution of
the data and by the s/n. High resolution data ($\simeq 25000-35000$)
with good s/n obtained at NTT by us \cite{gial96} showed that the
column density distribution is flatter than previously thought with a
slope $\sim 1.4$ for log~$N_{HI}<14$ after correcting for the
blanketing effects of the weak lines. At higher column densities the
distribution becomes steeper with a slope $\sim 1.8$. At the same time
the UVB value resulting from this distribution is $\sim J_{-22}=5\pm
1$ or $5_{-1}^{+2.5}$ if we allow for asymmetric errors. In this
analysis we have adopted QSO redshift estimates derived from low
ionization emission lines (OI, MgII, H$\alpha$) which are
representative of the true systemic redshift of the QSO. Recent Keck
data \cite{kim97} confirm the flat slope $\sim 1.4$ at lower column
densities.
Analyses derived from lower resolution and/or lower s/n
data assumed a single steep power-law distribution with $\beta
\sim 1.7$ resulting in a value of $J_{-22}\geq 10$ \cite{cooke97}.  
In summary, taking into account the correct bending of the column
density distribution and the often higher systemic QSO redshifts
provided by low ionization QSO emission lines we obtain a $J$ value
definitely lower than $10^{-21}$ and more similar to what expected
from the QSO contribution at $z\sim 2-3$ \cite{haard96}. Moreover there is no
indication of any appreciable redshift evolution of the UVB in the
redshift interval $z=1.7-4.1$ \cite{gial96}.

Of course this measure relies on the assumption that almost all of the
observed proximity effect is due to the increased photoionization
level near each QSO. However since high resolution spectra are taken
for the brightest high $z$ quasars, gravitational lensing can be
important. Indeed gravitational lensing can bias the estimate of the
ionizing UVB. If lensing brightens the QSO continuum then the QSOs are
intrinsically fainter than they appear, and $J$ is overestimated
\cite{bajt88}.  The fact that the deficiency of lines in QSOs of
different redshifts is correlated with luminosity but not with
redshift suggests that the statistical weight of the gravitational
lensing effect is small \cite{bechto94}.  However, comparison of the
proximity effect in lensed QSOs with unlensed objects would be an
interesting test of the photoionization model for the proximity
effect. A good example of this effect is shown by the high resolution
spectrum of the QSO 1208+10 (Fontana et al., these proceedings) where
the analysis of the proximity effect in its spectrum shows that the
QSO is lensed and at the same time that the proximity is strongly
dependent on the QSO luminosity, as expected from the photoionization
model of the proximity effect.

\section {The spectral shape of the UVB and the temperature of the Ly$\alpha$ 
clouds}

In the previous sections we have seen that the value $J_{-22}\simeq 5$
obtained from the photoionization model of the proximity effect is not
far from that predicted from the quasar contribution at
$z=3$. However, the absence of a strong redshift evolution in the UVB
implies that at $z\sim 4$ this agreement is not longer valid and a
discrepancy by a factor 3 does appear.
For this reason we have to look for possible different ionizing
sources like star forming galaxies.  Further information on the nature
of the ionizing sources comes in particular from the SiIV/CIV ratio
which is a sensitive function of the UVB shape between the HI and HeII
edge.  It has been shown that a QSO dominated UVB produces a $J$ ratio
between the HI and HeII edges $S_L\sim 30$ at $z=2-4$ \cite{haard96} while a
galaxy dominated UVB provides a value $S_L>300$ \cite{mir90}.
Thus a measure of the SiIV/CIV ratio can be translated in a measure of
this UVB ratio. We have obtained interesting measures from $z=2.8$ up
to $z=3.8$ for optically thin Lya clouds obtaining SiIV/CIV values in
the range 0.1-0.6. These large values imply a jump $S_L\geq 100$
specially at $z>3$ and confirm the increasing trend with increasing
$z$ already discussed by Cowie and Savaglio in these proceedings (but
see also the Boksenberg paper in these proceedings).
This trend suggests an evolution of the $S_L$ parameter that at $z\sim
2.5$ is consistent with that predicted from the quasar population
while at $z=3.5$ is more consistent with the value predicted by a
star-forming population.

It is interesting to note that a possible evolution of the UVB shape
can also explain the evolutionary trend observed in the minimum
temperature of the Lya clouds. Indeed Kim et al. \cite{kim97} found
that the lower envelope of the Lya doppler parameter, which is a
representative measure of the phtoionization temperature of the Lya
clouds, increases for decreasing redshifts, from b=15 at $z=3.7$ to
$b=24$ at $z=2.3$.
\begin{figure}[t]
\centerline{\vbox{ \psfig{figure=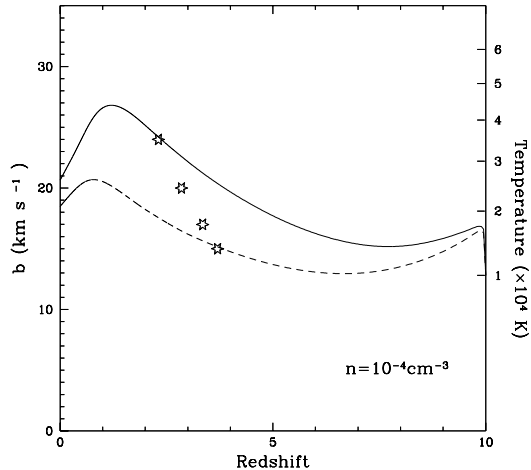,height=7.cm} }}
\caption[]{\label{fig1} Evolution of the Doppler parameter $b$ as a
function of redshift for a constant cloud density $n=10^{-4}$
cm$^{-3}$. Continuous curve: a power-law UVB with spectral
slope -1.5 is adopted; dashed curve: the same with a further jump by a
factor 100 at the HeII edge. Data points are from Kim et
al. \cite{kim97}.  }
\end{figure}
On the other hand, the presence of a strong HeII jump at the HeII edge
reduces the He contribution to the heating rate \cite{mira94,
GP94}. The low $b$ values and temperatures $T\sim 20000$K observed
at $z>3$ can be reproduced in large photoionized clouds.

Recently a new photoionization model has been proposed by Ferrara \&
Giallongo \cite{fer96} where Lya clouds are photoionized by a
time-dependent UVB, including non-equilibrium ionization effects. This
model shows that it is possible to account for low temperatures
(T=15000K or $b=15$) at $z\geq 3$ if the reionization epoch occurred
at $z\gg 5$ and the UVB has a jump by a factor 100 at the HeII edge. A
trend towards smaller $b$ with increasing redshift is present in the
redshift interval $z=1-5$ even for a fixed UVB shape because of the
cosmological evolution of the inverse compton cooling on the
CMB. However this evolution is not sufficient to explain that observed
and an evolution of the jump should be invoked from a value $\sim 100$
to e.g. $\sim 10$ for a cloud with $n\sim 10^{-4}$ (Fig.~1). Thus the
evolution in the UVB at the HeII edge drives both the observed
SiIV/CIV evolution and the temperature evolution of the weaker Lya
clouds.

\section {The Galaxy UV background and the redshift evolution of the
baryon density of photoionized gas}

In presence of a large UVB jump at the HeII edge, hot massive stars in
star forming galaxies could be considered as important contributors to
the UVB.  It has also been shown that the ionizing UV flux that
accompanies the production of metals at high-z can be comparable to
the QSO contribution if a fraction $\sim 25$\% of the UV radiation
emitted from stars can escape into the intergalactic space
\cite{mad96}.
On the other hand, at low and intermediate redshifts, the
Canada-France redshift survey \cite{lilly95} has provided new
information on the properties and evolution of field galaxies at
$z<1.3$.  The luminosity function of blue galaxies in the
Canada-France Redshift Survey shows appreciable evolution in the
redshift interval $z=0-1.3$, and generates a background intensity at 1
ryd of $J_L\approx 1.3\times 10^{-21}\langle f_{\rm esc}\rangle$ ergs
cm$^{-2}$ s$^{-1}$ Hz$^{-1}$ sr$^{-1}$ at $z\approx 0.5$, where
$\langle f_{\rm esc}\rangle$ is the unknown fraction of stellar
Lyman-continuum photons which can escape into the intergalactic space,
and we have assumed that the absorption is picket fence-type. It has
been argued that recent upper limits on the local UVB $\sim 8\times
10^{-23}$ (from the H$\alpha$ surface brightness of nearby
intergalactic clouds) constrain this fraction to be $\leq 20$\%
\cite{gial97}. If we adopt $\langle f_{\rm esc} \rangle \sim
15\%$ as a fiducial value for the escape fraction, the ensuing UVB is
found to evolve little in the redshift interval between $z=0.4$ and
$z=4$.
This small evolution of $J$ could have interesting consequences on the
evolution of the baryon fraction associated with the photoionized
Ly$\alpha$ clouds \cite{gial97}. Indeed the mass density parameter
depends on the volume filling factor times the average total volume
density of the clouds. Assuming photoionization equilibrium this
relation can be recast as a function of the typical observable
quantities like $J$, the size $l$ along the line-of-sight and the
number density of lines as a function of (z,NHI). It is possible to
notice that $\Omega$ depends only weakly on $J$ and on the typical
cloud size and geometry. The dominant factor is the number density
evolution of the clouds ($\Omega_{\rm IGM}\propto (JR)^{1/2} dN/dz$).

We can compute the line contribution to $\Omega$ integrating over the
known column densities and $z$ distributions. The main parameters
subjected to substancial uncertainties are the cloud sizes and
geometry. From the statistical coincidence of absorption lines in
closely separated quasar pairs, sizes $\sim 200$ kpc are derived at
$z\sim 2$ \cite{becht94}. Observations at lower redshifts show even
larger sizes $\sim 300$ kpc independently of the cloud structure and
geometry \cite{din95}. The mild evolution implied by these preliminary
measures is consistent with the expectations of the standard CDM
scenarios where the dominant effect appears to be the Hubble
expansion. Thus we have assumed that the typical cloud sizes increase
with cosmic time following the cosmological expansion.
\begin{figure}[t]
\centerline{\vbox{ \psfig{figure=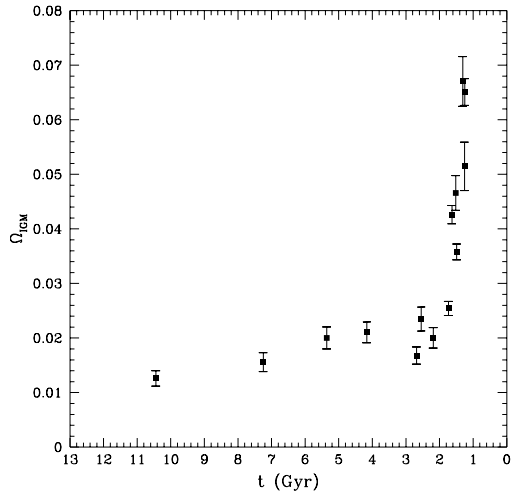,height=7.cm} }}
\caption[]{\label{fig2} Mass density parameter as a function of cosmic
time.  High $z$ points ($t<3$ Gyr) are derived from our optical
Ly$\alpha$ sample \cite{gial96}. Low $z$ points are derived from the
HST Ly$\alpha$ sample described in \cite{bach96}.

}
\end{figure}
We can see that at $z\sim 4$, $\Omega_{\rm IGM}$ may account for all
of the nucleosynthesis baryons of the universe for clouds with
thickness in the range $100-300$ kpc (Fig.~2).
Since $J$ remains constant between $z=2-4$, if the sizes do not
evolve faster than what expected from the Hubble expansion then the
significant evolution of $\Omega$ in this $z$ interval is mainly driven
by the evolution of the line number density.  At lower redshift the
evolution is mainly driven by the small evolution in the $(JR)^{1/2}$
factor since the observed number density is nearly constant as derived
by the HST observations \cite{bach96}.
A plausible explanation for the high $z$ evolution appears to be the
additional heating of the absorbing gas at temperatures $T\sim
10^{5-6}$ K by the gravitational accretion into progressively more
massive halos, with higher velocity dispersions, or by collisional
ionization from supernovae winds.  The easiest way for finding
collisionally ionized, cosmologically distributed material at $T\geq
10^{5.5}$ K is to look for OVI absorption.  The recent results of the
first survey for OVI 1032, 1038\AA\ absorption lines in QSO spectra
\cite{burles96} suggest the presence of a substantial cosmological
mass density of hot, collisionally ionized gas at $\langle z\rangle
=0.9$.  If the bulk heating were mainly due to supernovae explosions
in spheroidal systems, the strong evolution of $\Omega_{\rm IGM}$
observed between $z=2$ and $z=4$ could be triggered by the
star-formation activity in galaxies at high redshift.

\vfill
\end{document}